\documentclass{elsart}
\usepackage{graphics}
\usepackage{graphicx}
\usepackage{epsfig}
\usepackage{amssymb}

\begin{document}
\begin{frontmatter}
\title{Global Examination of the $^{12}$C+$^{12}$C Reaction Data at
Low and Intermediate Energies}
\author{Y. Kucuk and I. Boztosun}
\ead{boztosun@erciyes.edu.tr}
\address{Department of Physics, Erciyes University, 38039 Kayseri, Turkey}
\begin{abstract}
We examine the $^{12}$C+$^{12}$C elastic scattering over a wide
energy range from 32.0 to 70.7 MeV in the laboratory system within
the framework of the Optical model and the Coupled-Channels
formalism. The $^{12}$C+$^{12}$C system has been extensively studied
within and over this energy range in the past. These efforts have
been futile in determining the shape of the nuclear potential in the
low energy region and in describing the individual angular
distributions, single-angle 50$^{0}$ to 90$^{0}$ excitation
functions and reaction cross-section data simultaneously. In order
to address these problems systematically, we propose a potential
that belongs to a family other than the one used to describe higher
energy experimental data and show that it is possible to use it over
this wide energy range. This potential also predicts the resonances
at correct energies with reasonable widths.
\end{abstract}
\end{frontmatter}
\section{Introduction}
The $^{12}$C+$^{12}$C reaction has attracted enormous interest over
the years and considerable effort has been devoted to the
theoretical and experimental studies of this system. There is a
large body of experimental data measured so far
\cite{Bro60,Alm60,Erb81,Cor77,Cor78,Ful80,Cos75,Cos80,Sto77,Sto79,morsad,Kolata}
which have been attempted to be explained theoretically by using
both phenomenological and microscopic potentials (see
\cite{Bromley,Gre1,Sat97,freer04,Mar86,Flo84,Kon79,Abe80,Mat78,Mor91,Row77,Bra90,Kon98,mcvoy,Bra96,Boz1,michel}
for the details and the references therein).

In the last two decades, the scattering observables of this reaction
have been described by using the Optical model and Coupled-Channels
formalisms. Significant progress has been achieved in explaining the
interaction potential between two nuclei, in particular, in the
energy region of 6 MeV per nucleon and over. Their angular location
and cross-section have led to the determination of the gross
features of the local Optical potentials and ambiguities have been
clarified in many cases regarding the depths of the real parts of
the nuclear potentials \cite{Sat97}. A good understanding of the
theoretical basis of their features has been provided. As pointed
out in reference \cite{Sat97}, the resulting phenomenological
potentials are strongly attractive, with relatively weak absorption,
and they depend upon the bombarding energy.

However, a simultaneous description of the angular distributions,
excitation functions, resonances and total reaction cross-section
data have not been provided in a systematic way so far, for the
energies around 6 MeV per nucleon and under. For this energy region,
three different types of theoretical calculations have been
conducted:

The first type of calculations focuses on the observed resonances
\cite{Bro60,Alm60,Erb81,Cor77,Cor78,Ful80,Cos75,Cos80}, which are
one of the outstanding features of the light-heavy-ion collisions.
It has been speculated \cite{freer04} that these broad resonances
may be traced to either rotational structures
\cite{Mar86,Flo84,Ord86}, coupling of the elastic scattering channel
to the inelastic channel via the crossing of aligned molecular bands
(the band crossing model) \cite{Kon79,Abe80,Mat78}, or interference
effects resulting from reflected and refracted waves within the
nuclear medium \cite{Mor91,Row77}.

The second type of calculations concentrates on the excitation
functions and reaction cross-section data
\cite{Sto77,Sto79,morsad,Kolata,Bromley,Gre1,Sat97,Bra90,Kon98,mcvoy,Bra96,Boz1,michel}.
Among these studies, Brandan \emph{et al} \cite{Bra90} have extended
their potentials which fit the high experimental data ($E/A>6$) to
lower energies. From these analyses, mainly two types of deep
potentials have been obtained, called 'UNAM' potentials. UNAM
potentials explain the high energy experimental data in a systematic
way and they also predict the overall features of the 50$^{0}$ to
90$^{0}$ elastic scattering excitation functions and reaction
cross-section data. A slightly modified version of these potentials
has also been used in the excitation function analysis at low
energies \cite{Bra90,Kon98}. These two potentials have linear and
quadratic energy-dependence (Equation 6 and 7) in reference
\cite{Bra90}.

\begin{figure}[h]
\includegraphics[width=0.9\textwidth,angle=-90]{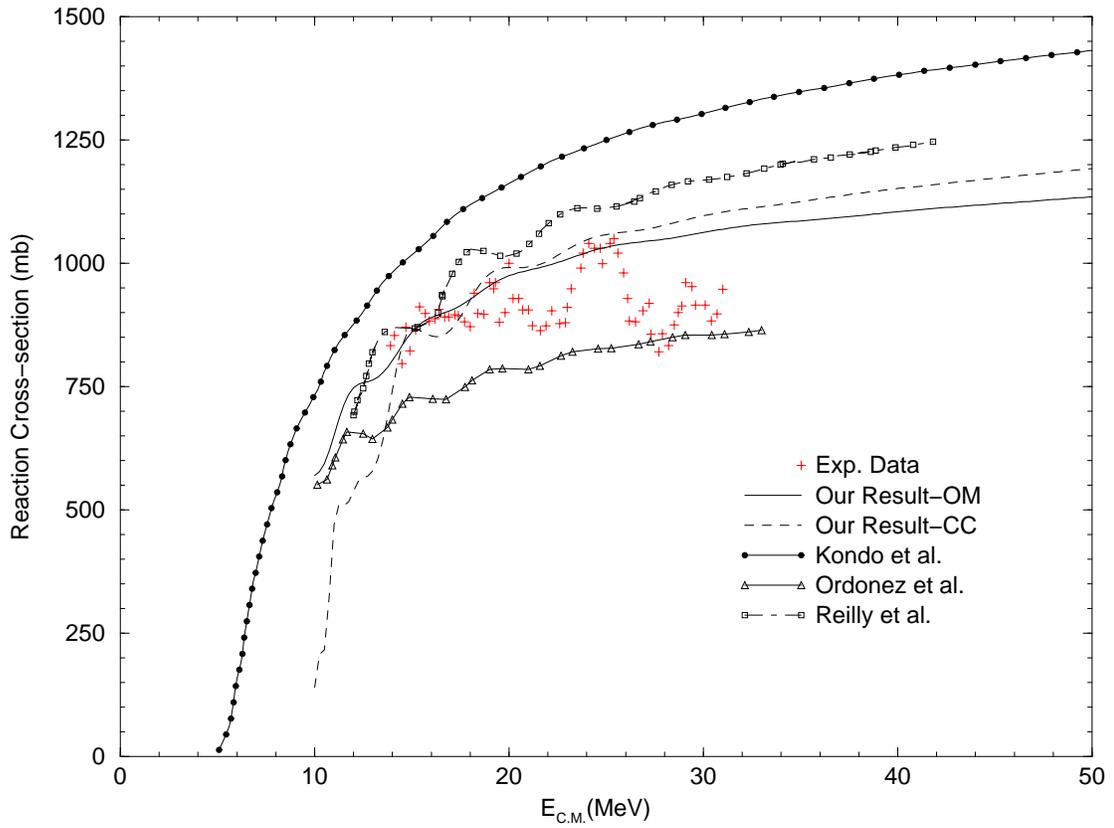} \vskip+1.0cm
\caption{Our reaction cross-section results by using Optical model
(solid lines) and Coupled-Channels model (dashed lines) are shown in
comparison with the experimental data and other theoretical
calculations conducted so far for the same experimental data. The
experimental data is taken from \cite{Kolata}.}
 \label{totalxsec}
\end{figure}

Although these potentials provide good agreement with the
experimental data for energies $E/A>6$, the analyses fail for the
excitation function and reaction cross-section data at low energies:
First of all, in the excitation function analysis, the
linearly-energy dependent potential (Equation 6) gives the correct
period and phase for the gross oscillations, but the strength is too
weak to reproduce the excitation functions at 60$^{0}$ and 80$^{0}$
(ref. \cite{Kon98}, Figure \ref{totalxsec}). The quadratic one
(Equation 7) gives a much worse prediction as shown in reference
\cite{Bra90}, the calculation exactly out of phase with respect to
the data for the 90$^{0}$ excitation function. An analysis similar
to the work of Brandan \emph{et al} \cite{Bra90} has also been
conducted by Kondo \emph{et al} \cite{Kon98}. They have modified the
UNAM potentials of Brandan \emph{et al} \cite{Bra90} even further in
order to explain 50$^{0}$ to 90$^{0}$ excitation function and
reaction cross-section data. Although there are still problems with
the strength and phases of the oscillation, an overall explanation
of the excitation functions and reaction cross-section data have
been provided by this work. Neither Brandan \emph{et al}
\cite{Bra90} nor Kondo \emph{et al} \cite{Kon98} shows in their
paper how good their predictions for the elastic scattering angular
distributions are.

Thirdly, Coupled-Channels calculations have been conducted for this
reaction. Boztosun and Rae \cite{Boz1} have recently analyzed this
reaction over a wide energy range from 32.0 to 126.7 MeV using a new
coupling potential. They have added to the usual first-derivative
coupling potential with a second-derivative one by considering the
orientation of two interacting nuclei and have attempted to solve
particularly the magnitude problem of the inelastic mutual-2$^{+}$
state cross-section at high energies. Their approach has provided a
good explanation of the elastic and inelastic angular distributions
as well as of the excitation functions and has solved the magnitude
problem for the inelastic mutual-2$^{+}$ state data, but there are
still justification problems regarding the use of such a coupling
potential.

In sum, the survey of the literature clearly shows that there is not
a potential model that can explain simultaneously the measured
elastic scattering angular distribution, the average behavior of the
50$^{0}$ to 90$^{0}$ excitation functions and reaction cross-section
data in the low energy region.

In the light of these studies, we have examined the experimental
data of $^{12}$C+$^{12}$C elastic scattering in low and intermediate
energy regions. The examination has been conducted in a systematic
way in order to find a potential family that simultaneously fits the
angular distributions and excitation functions data and to address
the problems of this reaction within the framework of the Optical
and Coupled-Channels models. 18 individual angular distribution, 5
excitation functions and reaction cross-section data between 32.0
and 70.7 MeV in the laboratory system have been studied, both within
the framework of the Optical model and the Coupled-Channels
formalism. The experimental data analyzed in this paper is taken
from \cite{Cos75,Cos80,Sto77,Sto79,morsad,Kolata}.

In the following section, we have introduced our Optical model as
well as the potential parameters. In this section, we have first
presented the results of the theoretical calculations for the
individual angular distributions, 50$^{0}$ to 90$^{0}$ single-angle
elastic scattering excitation functions and the reaction
cross-section data by using the Optical model and have compared them
with the experimental data. We have then provided the volume
integrals of the real and imaginary potentials and have discussed
their properties in terms of the dispersion relation. In Section
\ref{ccmodel}, we have also introduced our Coupled-Channels model
and the results of this analysis are shown for the individual
angular distributions, 50$^{0}$ to 90$^{0}$ single-angle elastic
scattering excitation functions and the reaction cross-section data.
The Optical model and Coupled-Channels calculations have been
compared and the effect of including the inelastic channels on the
scattering for low energies has been clearly demonstrated. Section
\ref{conc} is devoted to our summary and conclusion.

\section{Optical Model Calculations}
\label{model} In our Optical model analysis, we have used
phenomenological complex potentials:
\begin{equation}
V_{nuclear}(r) =
\frac{-V}{\left[1+e^{\frac{r-R_{V}}{a_{V}}}\right]^{2}} +i
\frac{-W}{1+e^{\frac{r-R_{W}}{a_{W}}}} \label{imagx}
\end{equation}

\begin{figure}[h]
\includegraphics[width=1.2\textwidth,height=0.9\textheight]{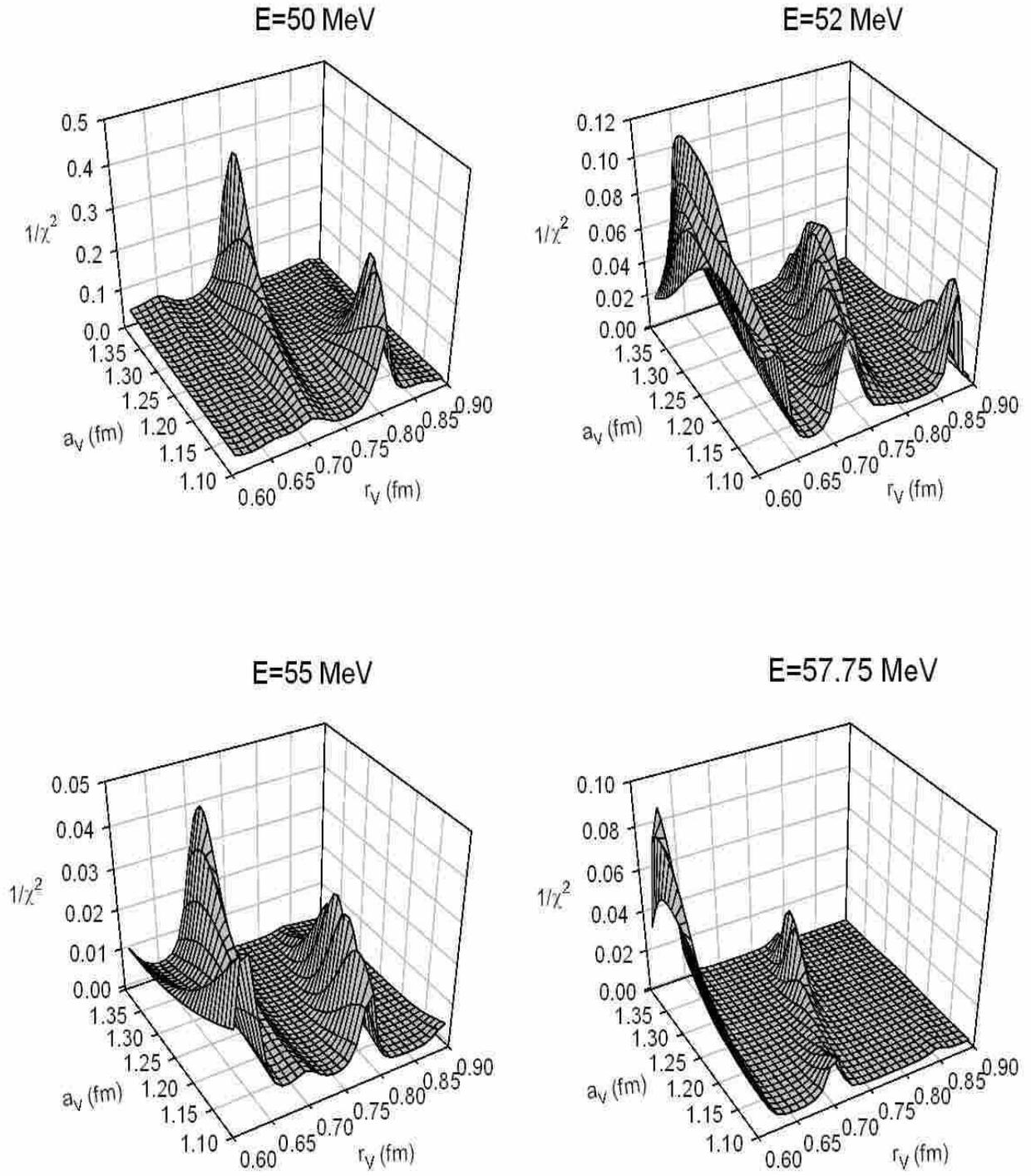}\vskip-0.0cm
\caption{Three-dimensional plots of the optical model parameters
$r_{V}$, $a_{V}$ versus 1/$\chi^{2}$, where $\chi^{2}$ has the usual
definition and measures the quality of the fit.} \label{chiall}
\end{figure}

\begin{figure}[h]
\includegraphics[width=1.0\textwidth]{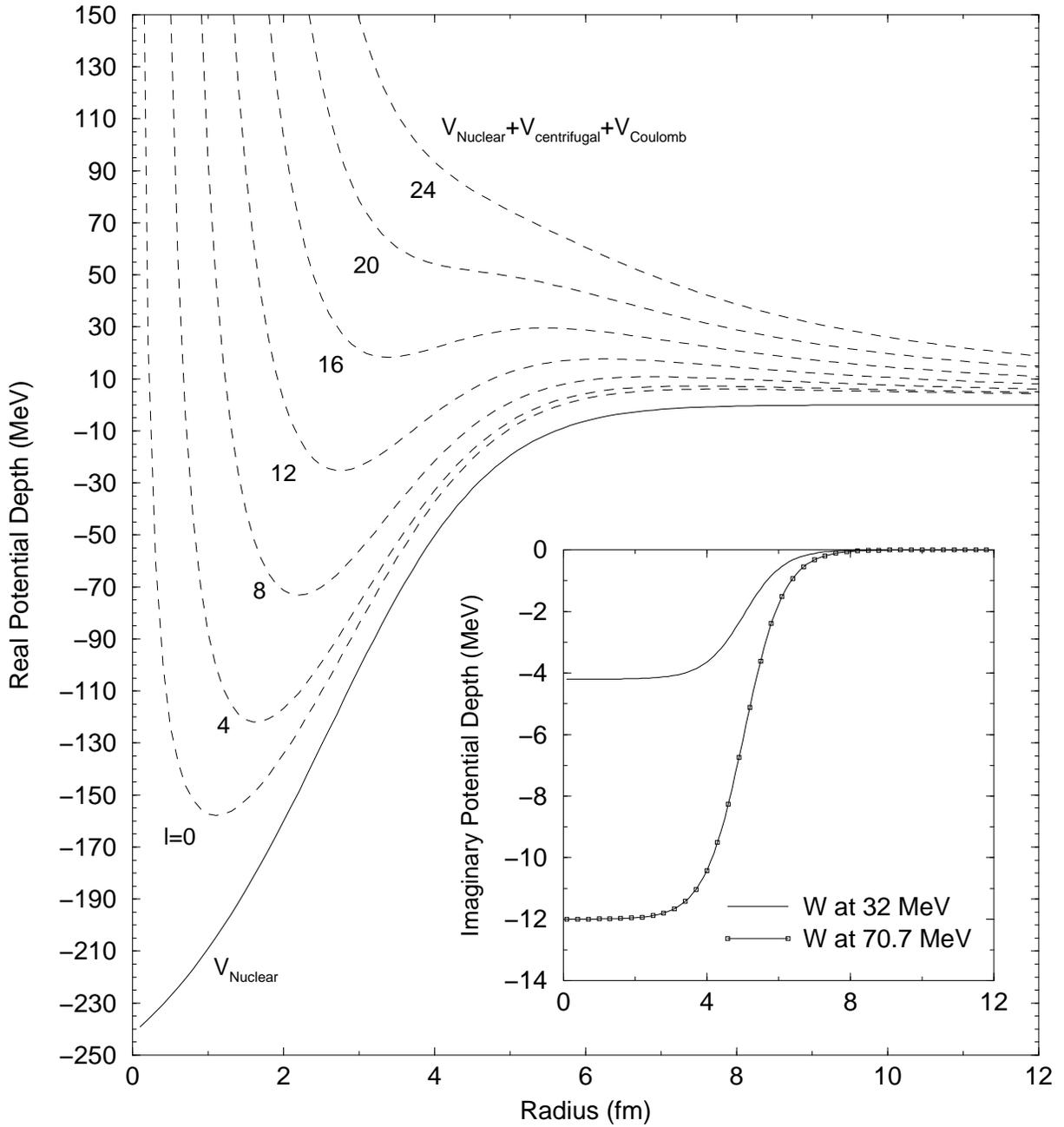}\vskip+0.0cm
\caption{The interaction potential between $^{12}$C and $^{12}$C is
plotted against the separation R for various values of the orbital
angular momentum quantum number, $l$. The inserted figure shows our
imaginary potential at $E_{Lab}$=32.0 MeV and 70.7 MeV. The
parameters are given in the text.} \label{realpot}
\end{figure}

Here, $R_{i}$=$r_{i} [A_{P}^{1/3}+A_{T}^{1/3}]$ ($i=V$ or $W$) where
$A_{P}^{1/3}$ and $A_{T}^{1/3}$ are the masses of projectile and
target nuclei and $r_{V}$ and $r_{W}$ are the radius parameters of
the real and imaginary parts of the nuclear potential respectively.
The real part of the nuclear potential has the square of the
Woods-Saxon shape and the depth ($V$=280.0 MeV) has been fixed to
reproduce the experimental data over the whole energy range
considered. The experimental data of the $^{12}$C+$^{12}$C reaction
in low energy region has a very oscillatory structure changing very
rapidly from energy to energy. Therefore, the radius ($r_{V}$) and
diffuseness ($a_{V}$) of the real potential have been varied on a
grid, respectively from 0.6 to 0.9 MeV, with steps of 0.01 MeV, and
from 1.1 to 1.5 fm with steps of 0.01 fm in order to obtain the best
fit to the data. The results of this systematic search are shown in
Figure \ref{chiall} which is a three-dimensional plot of the
$r_{V}$, $a_{V}$ and 1/$\chi^{2}$, where $\chi^{2}$ has the usual
definition and measures the quality of the fit. In Figure
\ref{chiall}, the best fit parameters, producing oscillating
cross-sections with reasonable phase and period, correspond to low
$\chi^{2}$ values and peaks in the 1/$\chi^{2}$ surface. For the
four incident energies, the figures present discrete peaks (or
hills) for correlated $r_{V}$ and $a_{V}$ values, which are best fit
real potential families and indicate that the $r_{V}$ or $a_{V}$
parameters cannot be varied continuously and still find equally
satisfying fits. For the radius ($r_{V}$), the lowest $\chi^{2}$
values are generally obtained between 0.72 and 0.78 and for the
diffuseness $a_{V}$, it is between 1.30 and 1.39. Therefore, the
radius ($r_{V}$) and diffuseness ($a_{V}$) of the real potential are
energy-dependent and the parameters are shown in Table \ref{param}.
The imaginary part of the nuclear potential given by Equation
\ref{imagx} has been taken as the Woods-Saxon volume form and its
depth (W) is given in Table \ref{param}. The other parameters of the
imaginary potential have also been fixed in the calculations as
$r_{W}$=1.1 fm and $a_{W}$=0.55 fm. The real and imaginary parts of
the nuclear potential are displayed in Figure \ref{realpot} for
various values of the orbital angular momentum. The Coulomb
potential is derived from a uniformly charged sphere with a radius
of 5.5 fm.

\subsection{Individual Angular Distributions}
\label{omresult1} 
\begin{figure}[h]
\includegraphics[width=1.0\textwidth]{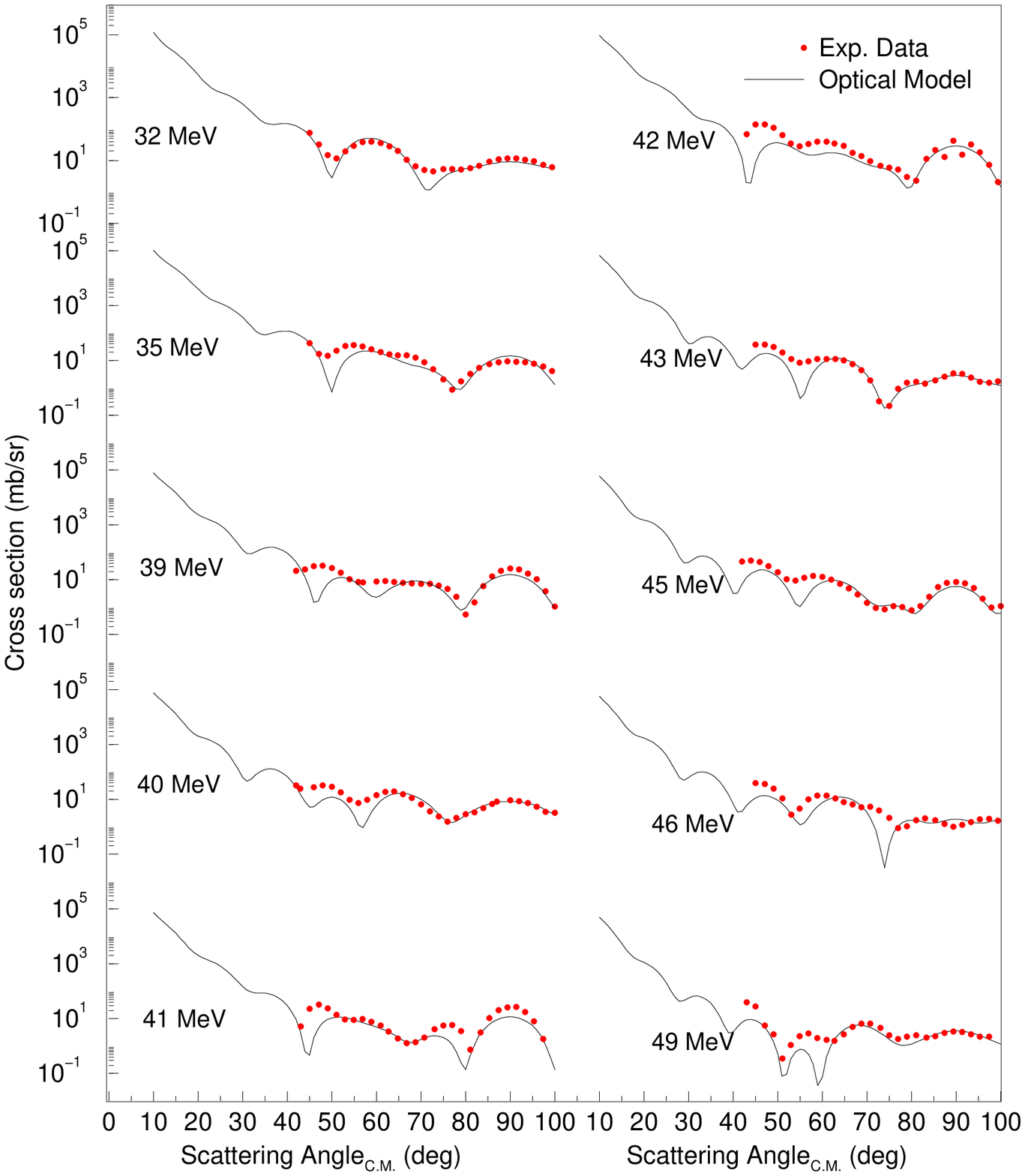} \vskip+1.0cm
\caption{The elastic scattering angular distributions obtained by
using the Optical model for the $^{12}$C+$^{12}$C system. The
experimental data is taken from \cite{Cos80}.} \label{ground1}
\end{figure}

\begin{figure}[h]
\includegraphics[width=1.0\textwidth]{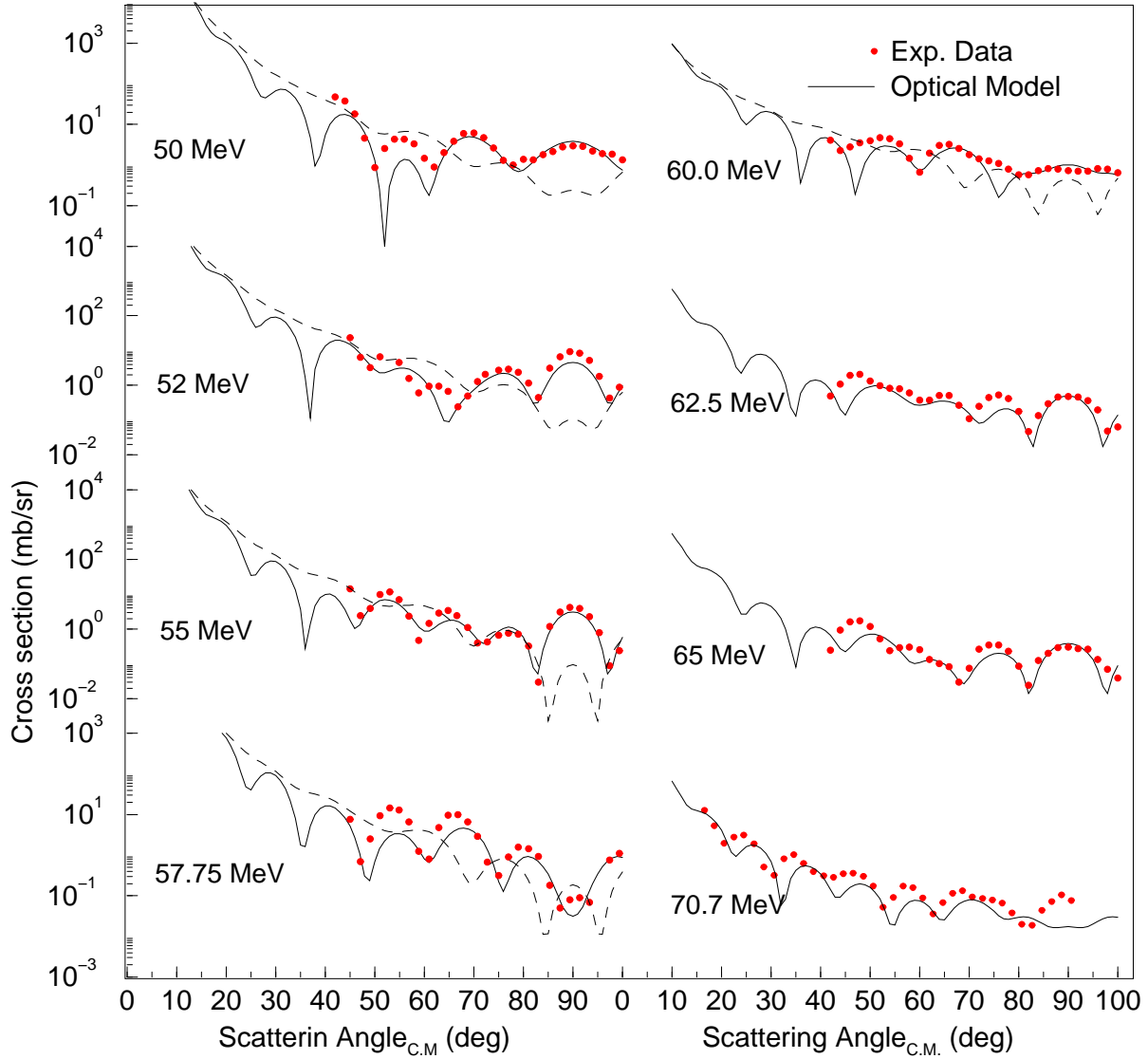} \vskip-2.0cm \caption{The elastic scattering angular
distributions obtained by using the Optical model for the
$^{12}$C+$^{12}$C system ({\it continued from Figure
\ref{ground1}}). The dashed lines are the angular distributions of
the UNAM potential by Brandan \emph{et al} \cite{Bra90}, which is
used to explain the excitation functions data at low energies. The
experimental data is taken from \cite{Cos80,Sto79}. }
\label{ground2}
\end{figure}

We have analyzed 18 angular distributions data
measured by references~\cite{Cos75,Cos80,Sto77,Sto79} between 32.0
and 70.7 MeV within the above-described Optical potential. The
parameters of the real and imaginary potentials as well as
$\chi^{2}$ values are given in Table \ref{param}. The volume
integrals of the potentials are also shown in Table \ref{param} and
they are displayed in Figure \ref{volint} in comparison with the
dispersion relation curve  between the real and imaginary parts of
the potential. The results of our Optical model calculations (solid
lines) are shown in Figures \ref{ground1} and \ref{ground2} in
comparison with the experimental angular distributions of
\cite{Cos75,Cos80,Sto77,Sto79} (circles). As it can be seen from
these figures, the places of the maxima and minima in experimental
data have been correctly reproduced and there is no magnitude
problem between our theoretical predictions and the experimental
data. Good agreement between the theoretical calculations and the
experimental data has been obtained within the framework of the
Optical model. We have also compared our Optical model results with
the results of the Brandan \emph{et al} \cite{Bra90} for 5 energies
between 50 and 60 MeV in figure \ref{ground2}. In this figure, the
dashed lines are the results of the UNAM potential by Brandan
\emph{et al} \cite{Bra90} which is used to explain the excitation
functions data at low energies. Although UNAM potential gives the
gross structure of the 90$^{0}$ excitation function, it fails to
give a good account of the individual angular distribution at low
energies. In the forward angle region up to almost 70$^{0}$ for some
energies, the theoretical cross-sections are structureless whereas
the experimental cross-section shows oscillatory structure.
\subsection{50$^{0}$ to 90$^{0}$ Excitation Functions}
\label{omresult2} It is pointed out in the introduction that the
theoretical calculations so far have had limited success in
describing the experimental data of this reaction. The potential
used to describe the individual angular distributions could not
predict the overall behavior of the excitation functions or
\emph{vice versa}. This may be seen from figure \ref{ground2} for
the elastic scattering angular distribution where the dashed-lines
are the predictions of UNAM potentials used to explain the
excitation functions and reaction cross-section data by Brandan
\cite{Bra90} and Kondo \emph{et al}  \cite{Kon98}. Therefore, we
have used our potential which fits the individual angular
distribution data in order to examine 50$^{0}$ to 90$^{0}$
excitation functions as well as reaction cross-section data. The
experimental data of references~\cite{Sto77,Sto79,morsad} for the
50$^{0}$, 60$^{0}$, 70$^{0}$, 80$^{0}$ and 90$^{0}$ excitation
functions have been analyzed using the parameters of
Table~\ref{param}. In the analysis of individual angular
distributions, we have used only 3 free parameters ($r_{V}$, $a_{V}$
and W) within the Optical model and we use now the linear
interpolation of the $r_{V}$, $a_{V}$ and W for the excitation
function calculations. The reason why we interpolate these
parameters is to prevent artificial peaks that might be created by
the change of parameters from energy to energy. The linear
interpolation of the radius, diffuseness and imaginary potential
parameters are
\begin{eqnarray}
r_V&=&0.80951-0.001226E_{Lab}  \nonumber \\
a_V&=&1.3482-0.00043225E_{Lab}  \nonumber \\
W&=&-0.77098+0.15341E_{Lab}\label{interp}
\end{eqnarray}

\begin{figure}[h]
\includegraphics[width=1.0\textwidth]{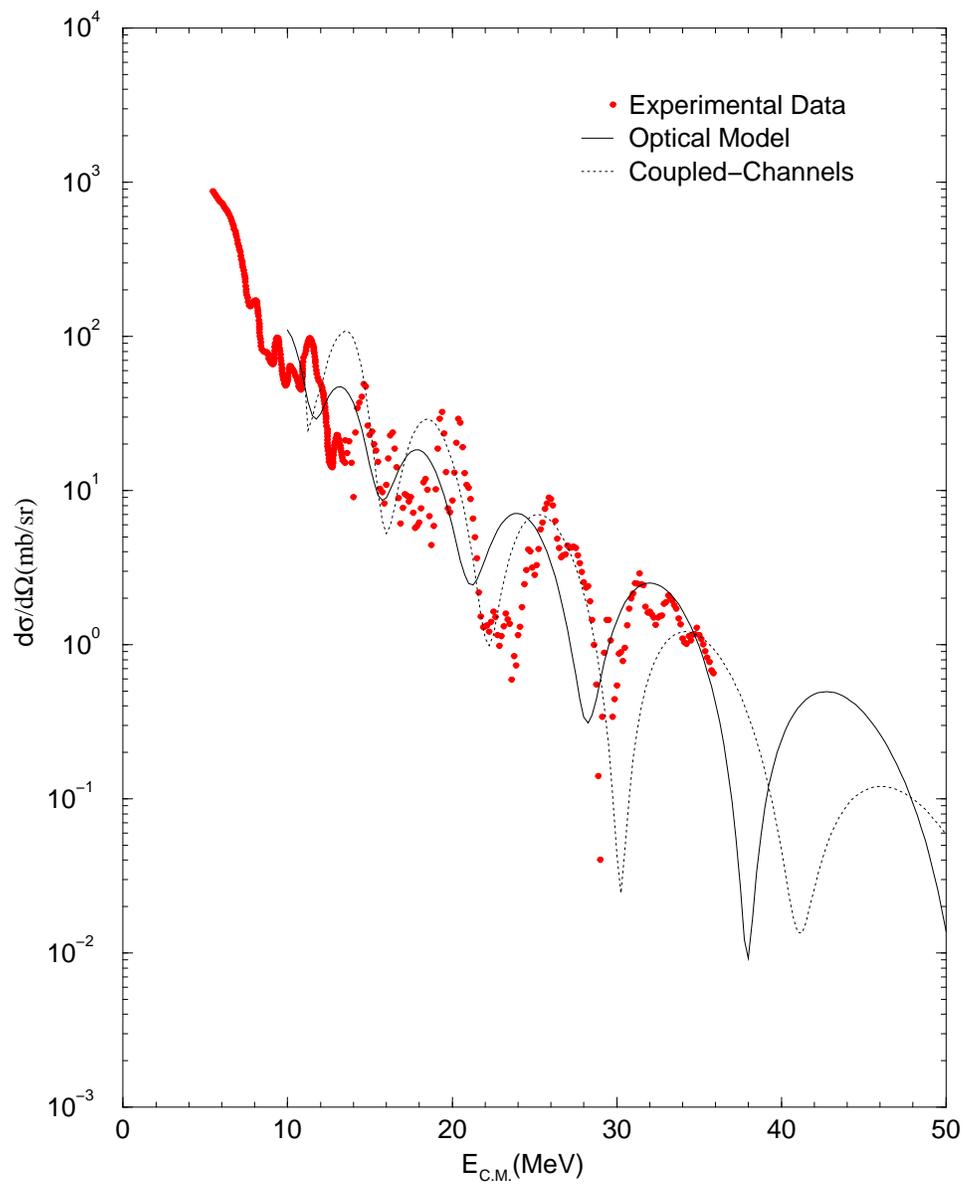} \vskip+0.0cm
\caption{The comparison of the excitation functions of the Optical
model prediction (solid lines) and Coupled-Channels prediction
(dashed lines) with the experimental data for the 90$^{0}$ elastic
scattering excitation function. The experimental data is taken from
\cite{Sto79,morsad}.}
 \label{exc-cc.ps}
\end{figure}

\begin{figure}[h]
\includegraphics[width=1.0\textwidth]{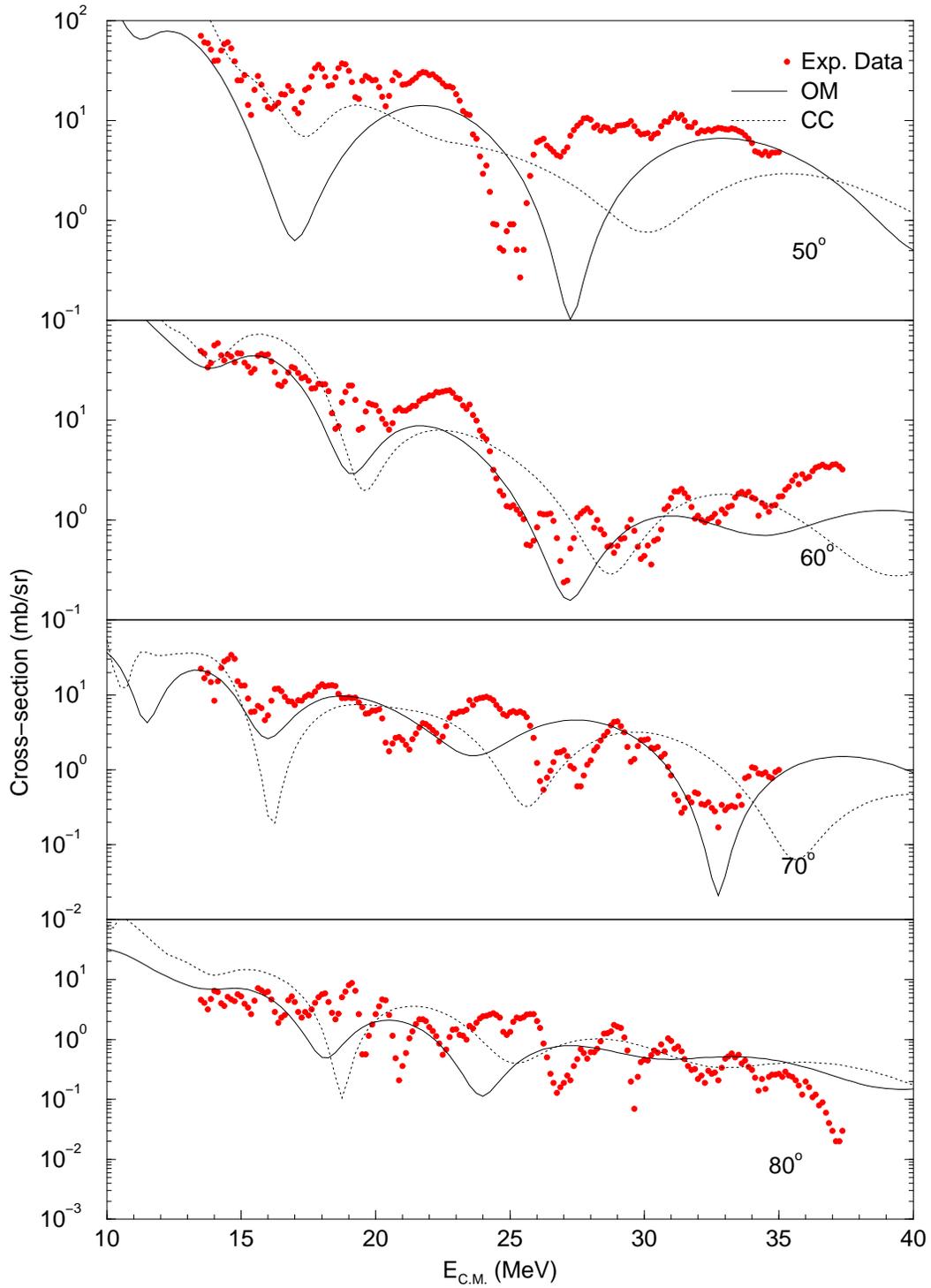} \vskip+0.0cm
\caption{The comparison of the 50$^{0}$, 60$^{0}$, 70$^{0}$ and
80$^{0}$ elastic scattering excitation function results with the
experimental data by using the Optical model (solid lines) and
Coupled-Channels model (dashed lines). The experimental data is
taken from \cite{morsad}.}
 \label{excnew.ps}
\end{figure}

The results of our Optical model calculations with these
interpolations are shown in Figure \ref{exc-cc.ps} for the 90$^{0}$
excitation function and in Figure \ref{excnew.ps} for the other
angles. It may be perceived from these figures and the $\chi^{2}$
values in table \ref{chisqaure} that the same Optical potentials
fitting the individual angular distributions could predict the
magnitude of the cross-section correctly and reproduce the period of
the gross-structure reasonably well.

\subsection{Reaction Cross-section Data}
\label{omresult3} In the same energy region, we have also examined
the reaction cross-section data of this reaction by using the same
potential parameters \emph{i.e.} the parameters in Table \ref{param}
and Equation \ref{interp} for the $r_{V}$, $a_{V}$ and W
interpolation. In Figure \ref{totalxsec}, our reaction cross-section
result is shown in comparison with the experimental data as well as
with other theoretical calculations conducted so far. We have
observed from this figure that our results are in very good
agreement with the experimental data; the magnitude and the phases
of the oscillations have been correctly provided by our theoretical
calculations up to $\sim$ E$_{Lab}$=60 MeV. It should be pointed out
that our calculations are in better agreement with the experimental
data in comparison with the other theoretical calculations conducted
so far for the same experimental measurement.

As a result, we have shown in this paper that a potential family
could explain 18 individual angular distributions data, 5 excitation
functions and reaction cross-sections data simultaneously in a
systematic way by using the Optical model. This outcome reveals that
our potential belongs to a different family other than the
potentials that have been previously used to analyze the
experimental data in the higher energy region \cite{Bra90,Kon98}.

\begin{figure}[h]
\includegraphics[width=1.0\textwidth]{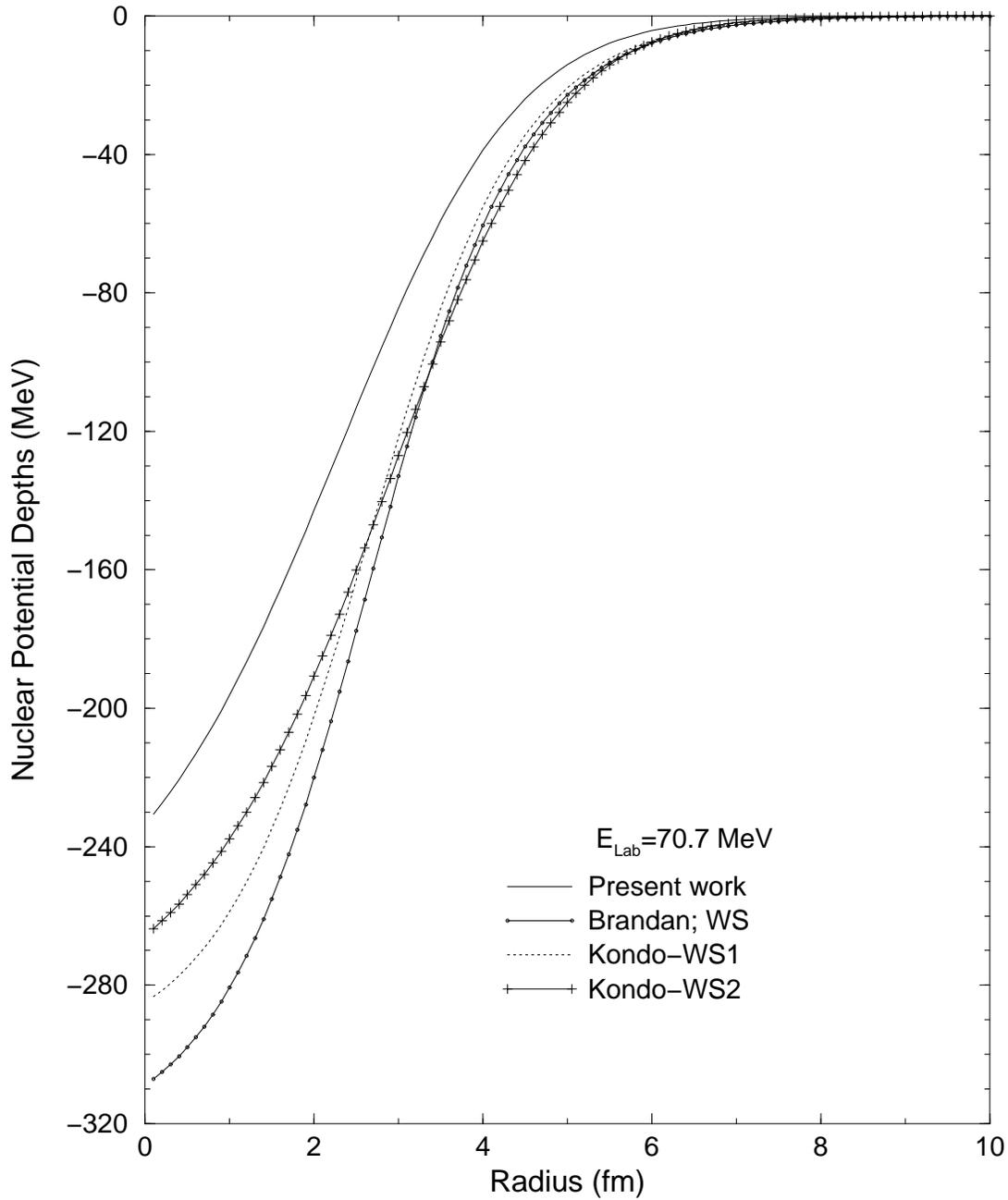}\vskip+0.0cm
\caption{The comparison of our nuclear potential, used in the
Optical model calculations, with the Brandan {\it et al}'s
\cite{Bra90} and Kondo {\it et al}'s \cite{Kon98} nuclear potentials
used in the analysis of the experimental data at high energies
(E/A$\geq$6).} \label{potcomp}
\end{figure}

In Figure \ref{potcomp}, we compare our real potential with the UNAM
potentials of Brandan {\it et al} \cite{Bra90} and Kondo {\it et al}
\cite{Kon98} which are successful in describing the high energy data
but fail to explain the low energy data. It may be perceived in this
figure that our real potential is more diffusive and shallower than
their potentials which is a phenomenon distinguishing it from other
potentials in the surface region where it is almost half as big as
other potentials. Similar differences may be asserted for the
imaginary potentials.

\begin{figure}[h]
\includegraphics[width=1.0\textwidth,height=1.0\textheight]{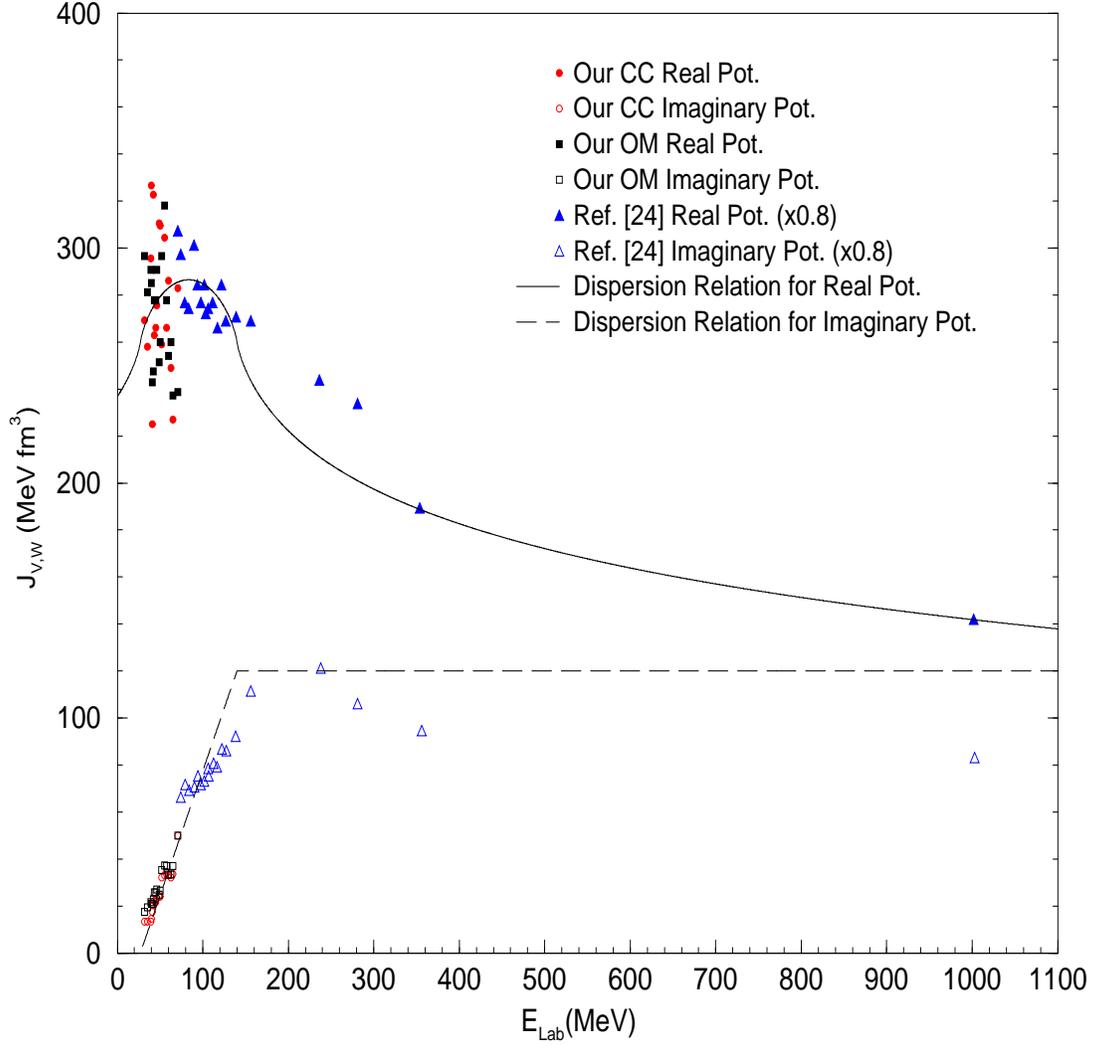}\vskip-4.0cm
\caption{The volume integrals of the real and imaginary parts of the
nuclear potential used in the Optical model (filled and empty
squares) and Coupled-Channels calculations (filled and empty
circles). Here, the filled and empty triangles are the volume
integrals of Brandan \emph{et al}'s potentials \cite{Bra90} and the
solid and dashed lines are the dispersion relation between real and
imaginary components of the nuclear potential calculated by using
equation \ref{disp}.} \label{volint}
\end{figure}

The volume integrals and the dispersion relation between the real
and imaginary potentials are shown in Figure \ref{volint} in
comparison with Brandan \emph{et al}'s potential \cite{Bra90}. The
volume integrals of the real and imaginary potentials and the
dispersion relation \cite{sat91} between them have been calculated
by using the following formula:
\begin{equation}
J_{V,W}(E)=-\frac{4 \pi}{A_{P}A_{T}}\int_{0}^{R}V,W(r,E)r^{2}dr
 \label{volintformula}
\end{equation}
\begin{equation}
V_{N}(E) = V_{R}+\triangle V(E) = V_{R}-(W/\pi)\left[\varepsilon_{a}
ln|\varepsilon_{a}|-\varepsilon_{b} ln |\varepsilon_{b}|\right]
 \label{disp}
\end{equation}
Here $\varepsilon_{i} = (E-E_{i})/(E_{b}-E_{a})$ with $i$ as $a$ and
$b$ respectively. The parameters are $E_{a}=32.0$ MeV, $E_{b}=140.0$
MeV, $V_{R}=260.0$ MeV and $W=120.0$ MeV.

We can derive from Figure \ref{volint} that the real potential does
not obey the dispersion relation at resonance energies observed by
Cormier {\it et al} \cite{Cor77,Cor78} for single and mutual-2$^{+}$
states for J=14, 16 and 18 spin values at E$_{Lab}$=$\sim38.0,
\sim50$ and $\sim57.5$ MeV respectively. In this resonance region,
the volume integral of the real potential oscillates remarkably, but
the imaginary potential does not accompany the variation of the real
one. One explanation of this may be due to the rapid variation of
the experimental data, which can not be described by a smoothly
varying parameter set and thus, the parameters oscillate. This is
also clearly seen from our $\chi^{2}$ search results in Figure
\ref{chiall}. For example, the lowest $\chi^{2}$ values for
E$_{lab}$=50 and 52 MeV in this figure require two very different
$r_{V}$ and $a_{V}$ values although there are only $2$ MeV energy
difference between two data. It is also possible to interpret that
the oscillations in the real potential are manifestations of the
coupling to the 2$^{+}$ state, due to the strongly deformed
structure of the $^{12}$C nucleus.

In a previous report \cite{Bra90,Boz1}, the strongly deformed
structure of the $^{12}$C nucleus has been examined and it has been
shown that Coupled-Channels calculations with a 20$\%$ decreased
imaginary potential of the Optical model calculations have provided
an equally good fit to the experimental data. This outcome has shown
that the inclusion of the inelastic channels mainly affects the
absorption and they have almost no effect in the real part of the
nuclear potential in the high energy region for the
$^{12}$C+$^{12}$C system.

\begin{figure}[h]
\includegraphics[width=1.0\textwidth]{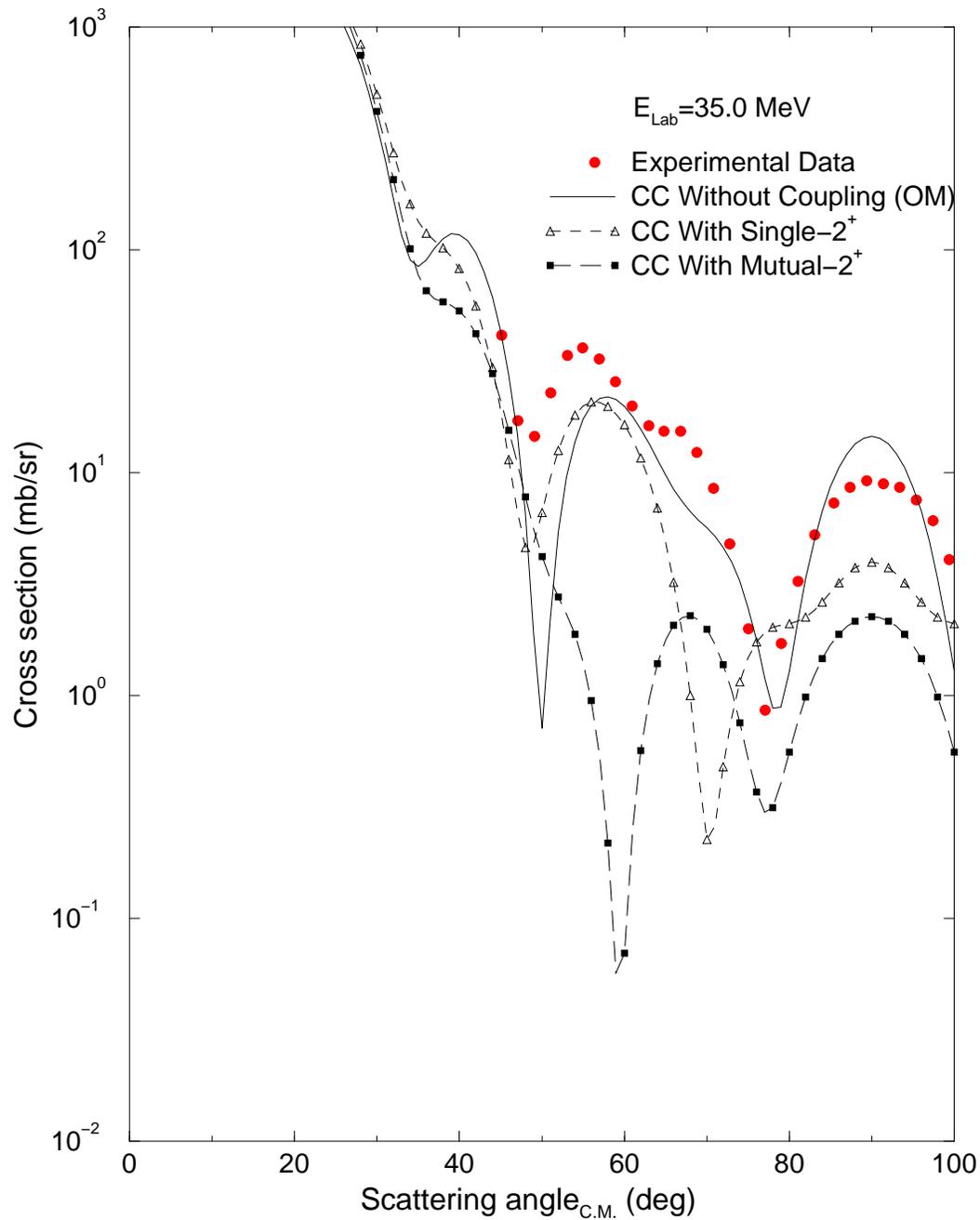} \vskip-0.0cm \caption{The effect of the inclusion of the
single-2$^{+}$ and mutual-2$^{+}$ excited states of the $^{12}$C
nucleus at E$_{Lab}$=35.0 MeV. The Optical model parameters of Table
\ref{param} are used in the calculations. See the text for the
discussion.}
 \label{ccresults1}
\end{figure}

However, other previous works such as Sakuragi and Kamimura
\cite{Sak84} and references therein have investigated the breakup
effect and have shown that the inclusion of the inelastic channels
induces a large repulsive real potential. The energies we have
studied in this paper are lower than the analysis of ref.
\cite{Bra90,Boz1} and the nature of the experimental data is very
different than the nature of those in higher energies. Therefore, it
should be questioned how the inclusion of the inelastic channels
affects the scattering observables of this reaction. For this
purpose, test calculations have been performed at E$_{Lab}$=35.0 MeV
to see the coupling effect for low energies. In these calculations,
we have included the single-2$^{+}$ (4.44 MeV) and mutual-2$^{+}$
(8.88 MeV) states of the strongly deformed $^{12}$C nucleus. The
results of test calculations obtained by using the Optical model
parameters as shown in table \ref{param} for E$_{Lab}$=35.0 MeV are
displayed in Figure \ref{ccresults1}.

Test runs  with and without coupling (\emph{i.e.} Optical model) at
E$_{Lab}$=35.0 MeV have, however, shown striking differences. As
shown in Figure \ref{ccresults1}, the inclusion of the excited
states at E$_{Lab}$=35.0 MeV has very large effects not only in
backward, but also in forward-angle regions
($\theta_{C.M.}\leq40^{0}$). The coupling changes the magnitudes and
phases of the oscillations in the cross-section. We have infered
from this effect that the coupling not only affects the absorptive
or imaginary part, but also the real part of the nuclear potential.
As a result, it modifies the interference between incoming and
outgoing waves, which creates the oscillatory structure in the
cross-section. Our findings are in agreement with the observation of
Sakuragi and Kamimura \cite{Sak84}. Therefore, according to this
outcome, the claim of ref. \cite{Bra90} that is the inclusion of the
inelastic channels mainly affects the absorption and they have
almost no effect in the real part of the nuclear potential in the
high energies is not valid for the low energy region the
$^{12}$C+$^{12}$C system.


\section{Coupled-Channels Calculations}
\label{ccmodel} Having seen the coupling effect, we conduct
Coupled-Channels calculation for the same energies considered in
Optical model calculations. In the Coupled-Channels calculations,
the interaction between the $^{12}$C nuclei is described by a
deformed Optical potential. The real potential has the square of a
Woods-Saxon shape as in Equation \ref{imagx} with a depth of 280
MeV. The other parameters, as shown in Table \ref{paramcc}, have
been fitted to reproduce the elastic scattering data. The imaginary
potential has the standard Woods-Saxon volume shape as in Equation
\ref{imagx} and the parameters of its depth are given in Table
\ref{paramcc}. The radius and diffuseness of the imaginary potential
have been fixed in the calculations as $r_{W}$=1.1 fm and
$a_{W}$=0.55 fm.

It has been assumed that the target nucleus $^{12}$C has a static
quadrupole deformation and this assumption has been taken into
account by deforming the real potential in the following way:
\begin{equation}
R(\theta,\phi)=r_{0}A_{p}^{1/3}[1+\beta_{2}
Y_{20}(\theta,\phi)]+r_{0}A_{t}^{1/3}[1+\beta_{2}
Y_{20}(\theta,\phi)] \label{beta}
\end{equation}
where the first and second terms account for the projectile and
target excitations respectively. In equation \ref{beta},
$\beta_{2}$=-0.6 is the deformation parameter of the $^{12}$C
nucleus. This value is derived from its known B(E2) value, which is
42 e$^{2}fm^{4}$ \cite{Ste66}. We have noticed that deformation of
the imaginary part of the nuclear potential does not have any
significant effect. Thus, for computational simplicity, we have not
deformed it in the present Coupled-Channels calculations. An
extensively modified version of the code Chuck and the code Fresco
has been used both in the Optical model and Coupled-Channels
calculations \cite{chuck,fresco1}.

\subsection{Individual Angular Distributions}
\label{ccresult1} Using this Coupled-Channels model, we have
analyzed the same experimental data as in the case of the Optical
model. The parameters of the real and imaginary potentials as well
as their volume integrals are given in Table \ref{paramcc}. The
volume integrals are also displayed in figure \ref{volint} in
comparison with the volume integrals of the Optical model potentials
and the dispersion relation curve obtained by using equation
\ref{disp}.

\begin{figure}[h]
\includegraphics[width=1.0\textwidth]{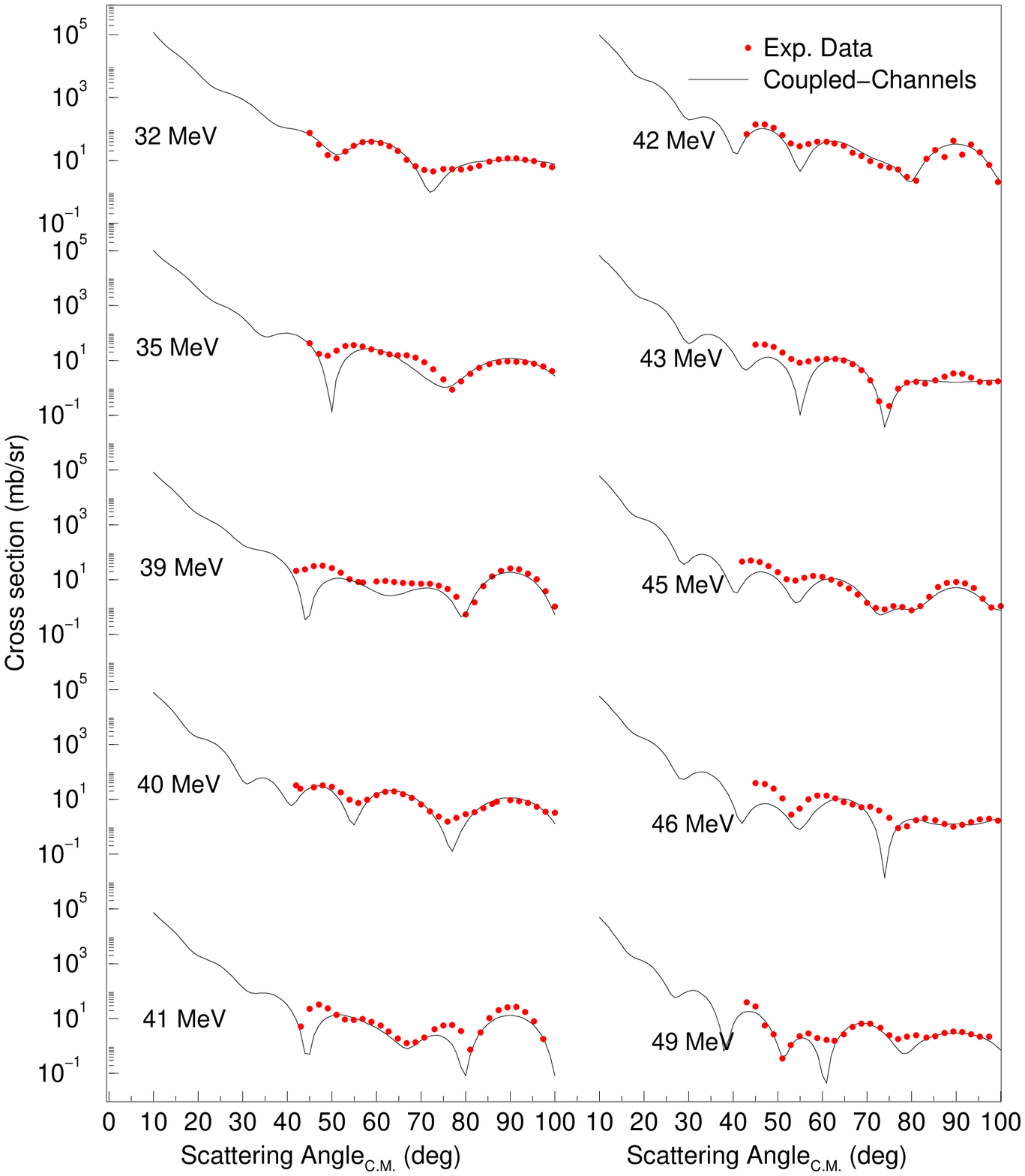} \vskip+1.0cm
\caption{The elastic scattering angular distributions obtained by
using the Coupled-Channels model for the $^{12}$C+$^{12}$C system.
The experimental data is taken from \cite{Cos80}.} \label{ground1cc}
\end{figure}

\begin{figure}[h]
\includegraphics[width=1.0\textwidth]{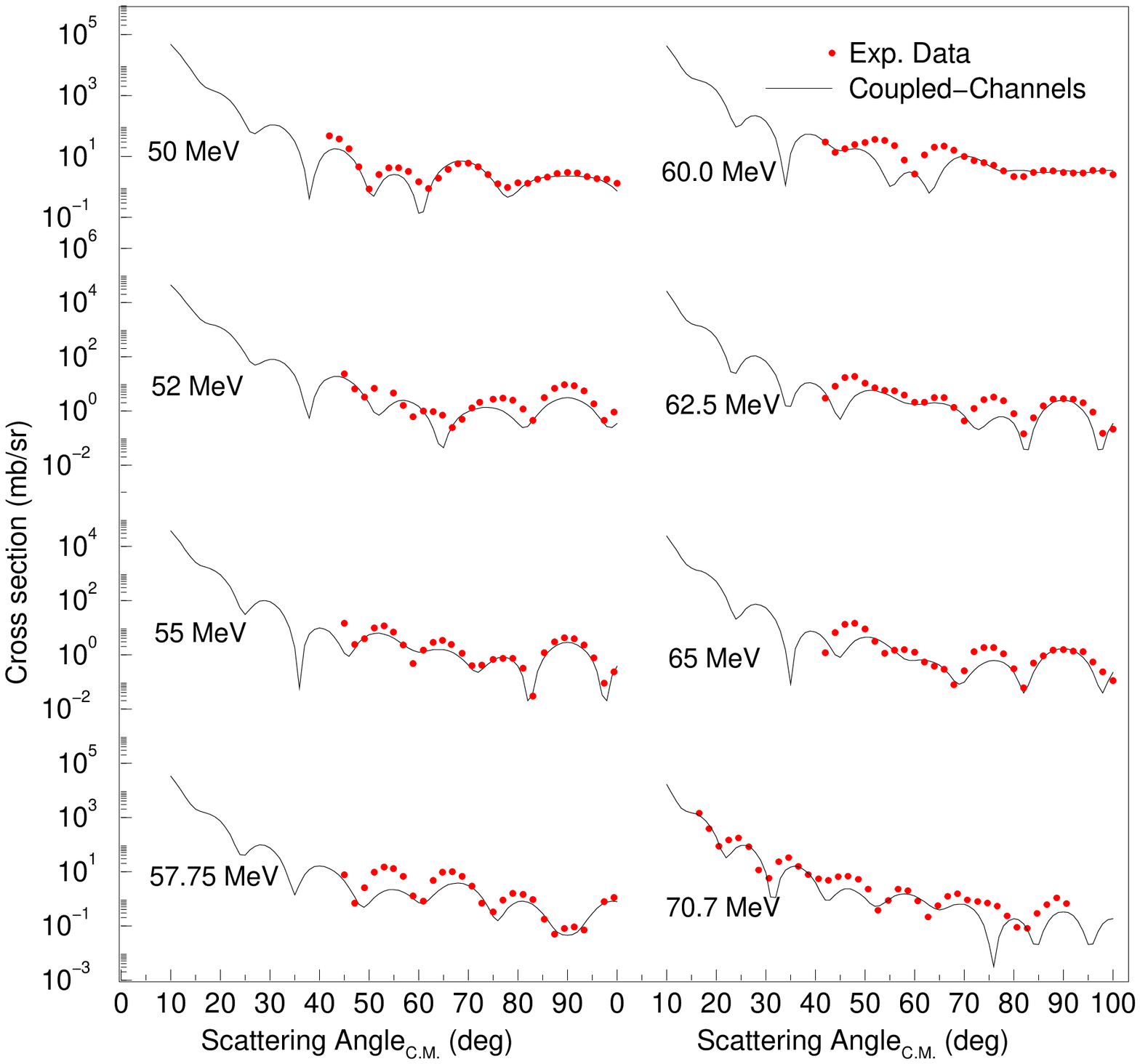} \vskip-2.0cm \caption{The elastic scattering angular
distributions obtained by using the Coupled-Channels model for the
$^{12}$C+$^{12}$C system ({\it continued from Figure
\ref{ground1cc}}). The experimental data is taken from
\cite{Cos80,Sto79}. } \label{ground2cc}
\end{figure}

The results of our Coupled-Channels calculations (solid lines) are
shown in Figures \ref{ground1cc} and \ref{ground2cc} in comparison
with the experimental data (circles). As it can be seen from these
figures and the $\chi^{2}$ values in Table \ref{paramcc}, reasonable
agreement between the theoretical calculations and the experimental
data has been obtained. The phases and magnitudes of the
oscillations have been predicted correctly for the energies
considered.

The effect of the inclusion of the single-2$^{+}$ state of the
$^{12}$C nucleus in Coupled-Channels calculations should be
underlined here. The effect of the Coupled-Channels calculations has
been very large on the scattering and its effect has been clearly
observed in our theoretical calculations for the cross-section at
forward, intermediate and large angles. It has changed the phases
and magnitudes of the oscillations at all angles. In order to obtain
the agreement between theoretical calculations and experimental
data, we have had to change not only the imaginary part, but also
the real part of the nuclear potential. As seen in figure
\ref{volint}, the inclusion of the single-2$^{+}$ state has reduced
the strength of the imaginary potential at all energies in
comparison with the imaginary potential of the Optical model. This
reduction of the imaginary potential was expected since the
Coupled-Channels calculations take into account the effect of the
eliminated channels of the Optical model. However, the coupling has
had a big effect on the real potential and has changed the strength
of the real potential very much. This remarkable change of the real
potential parameters has been due to the coupling between ground and
single-2$^{+}$ states and in contrast with the observations at high
energies. At high energies, the coupling does not change the phase
of the oscillations and therefore, reducing the imaginary potential
would be enough to obtain Optical model results. We should state
that this strong coupling effect should be taken into account in the
analysis in order to gain a better description of the nuclear
potential in the interaction of such strongly deformed two nuclei.

\subsection{50$^{0}$ to 90$^{0}$ Excitation Functions and Reaction Cross-section Data}
\label{ccresult2} We have also analyzed the 50$^{0}$ to 90$^{0}$
excitation functions as well as reaction cross-section data. We have
used linear interpolation of 3 free parameters given in Table
\ref{paramcc}, \emph{i.e.} $r_{V}$, $a_{V}$ and W, as follows:
\begin{eqnarray}
r_V&=&0.78664-0.00056643E_{Lab}  \nonumber \\
a_V&=&1.3476+0.00073848E_{Lab}  \nonumber \\
W&=&-4.1719+0.209E_{Lab} \label{interp2}
\end{eqnarray}

The results for the 90$^{0}$ excitation function are shown in Figure
\ref{exc-cc.ps}, for the other angles in Figure \ref{excnew.ps}, and
for the reaction cross-section data in Figure \ref{totalxsec} in
comparison with the Optical model results and experimental data.
Again reasonable agreement with the experimental data both for the
excitation functions and for the reaction cross-section data has
been obtained within the framework of the Coupled-Channels method.
However, we should point out that the Optical model results for the
excitation function calculations are in better agreement with the
experimental data as it can be seen from the $\chi^{2}$ values in
Table \ref{chisqaure}. For the reaction cross-section, the
Coupled-Channels results are better than the Optical model results
at low energies, but towards high energies our Coupled-Channels
calculations have over-estimated the data. It is higher than the
experimental data and Optical model results. This discrepancy arises
due to the interpolation formulae given in equation \ref{interp2}
for $r_{V}$, $a_{V}$ and W. Second or third-order interpolations of
the parameters in Table \ref{paramcc} have provided a better result
for the excitation functions and reaction cross-section data.

\subsection{Resonances}
\begin{figure}[h]
\includegraphics[width=1.0\textwidth]{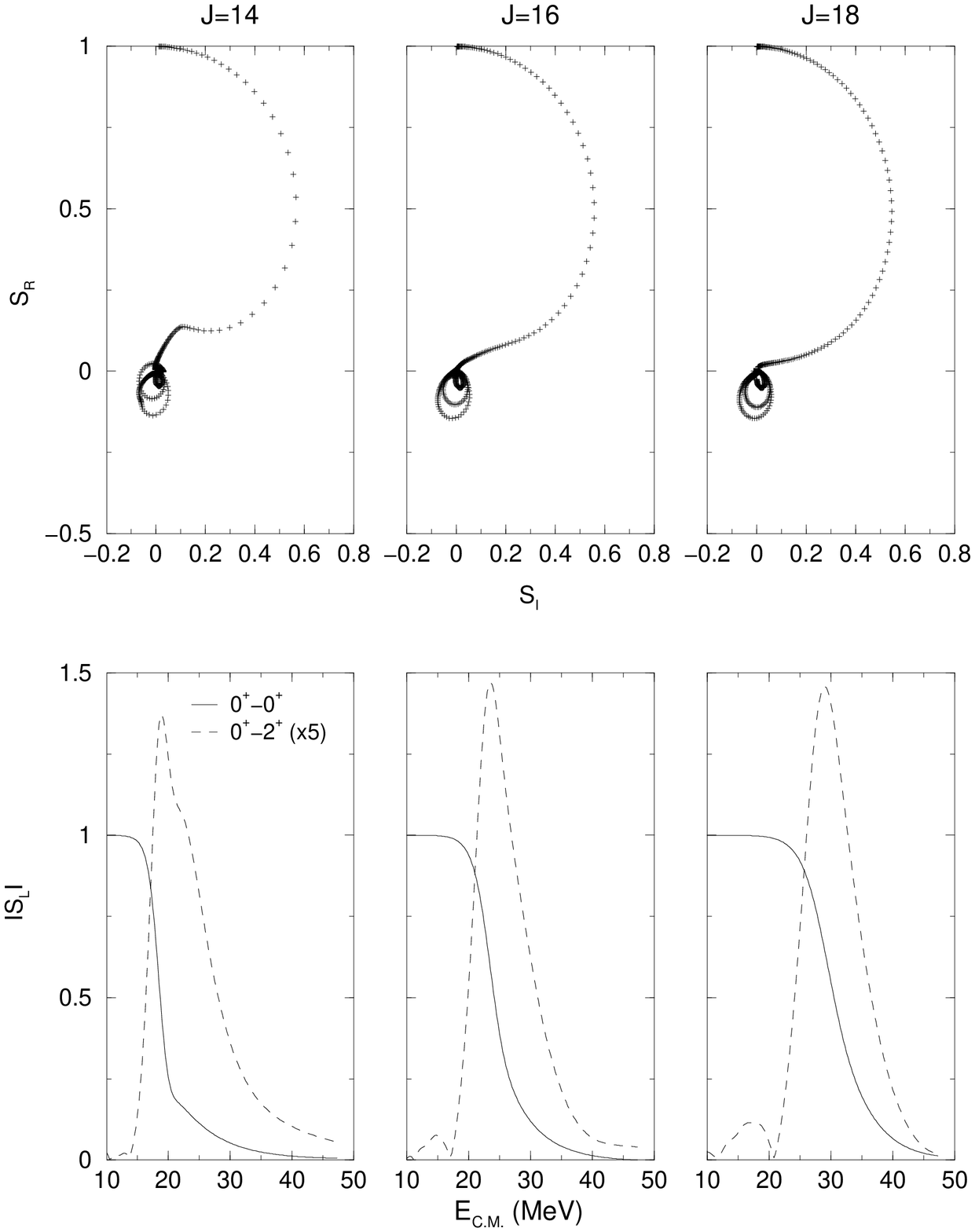} \vskip-0.0cm
\caption{Resonances: In the upper part of the figure, we show the
real versus imaginary part of the S-Matrix for the spin values J=14,
J=16 and J=18 and in the lower part, the magnitudes of the S-matrix
($|S_{L}|$) for the elastic ($0^{+}-0^{+}$) and single-$2^{+}$
($0^{+}-2^{+}$) channels against the center of mass energy for the
same spin values.} \label{smat}
\end{figure}

In this section, we give the prediction of our potential for the
well-known resonances measured by Cormier \emph{et al.}
\cite{Cor77,Cor78} for the single-2$^{+}$ state. These resonances
observed at low energies could not be predicted by a potential,
which also fits the angular distributions and the excitation
functions. We show the predictions of our potential in Figure
\ref{smat}. In the upper part of this figure, we display the real
versus the imaginary part of the S-matrix for the resonance spin
values J=14, J=16 and J=18 and in the lower part of the figure we
show the magnitudes of the S-matrix ($|S_{L}|$) for the elastic
($0^{+}-0^{+}$) and single-$2^{+}$ ($0^{+}-2^{+}$) channels against
the center of mass energy for the same spin values. Our potential
predict the places of the resonances at the correct energies with
reasonable widths.

\section{Summary and Conclusion}
\label{conc}

We provide a consistent description of the elastic scattering of the
$^{12}$C+$^{12}$C system from 32.0 to 70.7 MeV in the laboratory
system by using a phenomenological, strongly attractive Woods-Saxon
squared nuclear potential, with a relatively weak absorption, with
both the Optical model and Coupled-Channels formalism. This reaction
has been one of the most extensively studied reaction over the last
forty years, in particular, in the low energy region where an
oscillatory structure in the excitation functions and resonances
observed has been pronounced. Unfortunately, no global model has
been set forth so far that consistently explains the measured
experimental data over a wide energy range. In the introduction, we
present the problems that this reaction manifests, one of the most
important of which is the simultaneous description of the elastic
scattering angular distributions and the single-angle excitation
functions data. The theoretical calculations in the past have
reported that a potential family that fits the individual angular
distributions is unable to reproduce the excitation functions data
of the same reaction in a systematic way.

By considering these problems, we analyze this reaction with
available experimental data in the energy range we considered and
show that it is possible to improve the agreement for the individual
angular distributions, excitation functions and the reaction
cross-section data simultaneously with a deep real and shallow
imaginary potential. From our analysis, we observe that away from
the resonance energies, the variation of the potential parameters
has a systematic energy-dependence as in the case of the high energy
region. However, when there is a rapid variation of the experimental
data, it can not be described by a smoothly varying parameter set
and thus, the parameters oscillate. In our analysis, we observe this
effect. It should be underlined here that the remarkable success of
our potential in explaining the individual angular distributions,
excitation functions and the reaction cross-section data
simultaneously is achieved by using only 3 free parameters ($r_V$,
$a_V$ and $W$) in the calculations.

We should also point out that the coupling between ground and
excited states  has a large effect on the scattering in the low
energy region. The inclusion of the excited states of the $^{12}$C
nucleus at E$_{Lab}$=35.0 MeV  alters the places and magnitudes of
oscillations at both forward and backward angles around 90$^{0}$ in
the theoretical results. That means, the coupling not only changes
the absorptive part of the nuclear potential, but also the real part
of the nuclear potential which is in agreement with the microscopic
calculations of Sakuragi and Kamimura \cite{Sak84}. Our results show
that at low energies, the Coupled-Channels effect should be
explicitly taken into account for strongly deformed nuclei such as
$^{12}$C. At high energies, we observe that the inclusion of the
excited states does not change the phases of the oscillation but
affects the magnitude of the cross-section at large angles. We do
not observe effects at forward angles. No effect on the phases of
the oscillations means that the inclusion of the excited states has
almost no effect on the real part of the nuclear potential. The
effect at large angles can also be compensated by decreasing the
absorptive part of the nuclear potential. This outcome is in
agreement with previous claims \cite{Bra90,Boz1}, but valid only at
high energies.

The deep real potential used in our Optical and Coupled-Channels
analysis has a different radial shape, separating it from other
potentials used in the analysis of the higher energy data
($E/A\geq6$ MeV). Both Optical and Coupled-Channels potentials in
our analysis  provide very good fits to the elastic scattering
angular distributions and reasonable improvement to the 50$^{0}$ to
90$^{0}$ excitation functions and reaction cross-section data. It
also predicts the resonances at the correct energies with reasonable
widths. To our knowledge, this has not been achieved for the
$^{12}$C+ $^{12}$C reaction at low energies so far.


This work is supported by the Turkish Science and Research Council
(T\"{U}B\.{I}TAK), Grant No: TBAG-2398 and Erciyes
University-Institute of Science: Grant no: FBA-03-27, FBT-04-15,
FBT-04-16.
%

\newpage
\begin{table}[h]
\begin{center}
\begin{tabular}{lccccccc} \hline
$E_{Lab}$ & $r_{V}$ &$a_{V}$&$W$ &$J_{V}$ &$J_{W}$& $\chi^{2}$ \\
MeV & fm & fm & MeV & MeV.fm$^{3}$ & MeV.fm$^{3}$ &  \\ \hline
32.0  &0.78  &1.39 &  4.2   &  296.5  &  17.4   &  2.51 \\
35.0  &0.77  &1.33 &  4.7   &  281.2  &  19.5   &  4.74 \\
39.0  &0.78  &1.33 &  5.2   &  290.7  &  21.6   &  6.32 \\
40.0  &0.77  &1.37 &  5.0   &  285.0  &  20.8   &  4.19  \\
41.0  &0.72  &1.39 &  5.0   &  242.9  &  20.8   &  5.62 \\
42.0  &0.73  &1.35 &  5.5   &  247.5  &  22.8   &  7.72  \\
43.0  &0.76  &1.39 &  6.2   &  277.7  &  25.8   &  3.62  \\
45.0  &0.76  &1.39 &  6.2   &  277.7  &  25.8   &  4.59  \\
46.0  &0.78  &1.33 &  6.5   &  290.7  &  27.0   &  6.36    \\
49.0  &0.73  &1.39 &  6.0   &  251.3  &   24.9  &  5.58  \\
50.0  &0.74  &1.39 &  6.4   &  259.9  &  26.6   &  4.95   \\
52.0  &0.78  &1.39 &  8.5   &  296.5  &  35.3   &  9.88   \\
55.0  &0.81  &1.30 &  9.0   &  318.1  &  37.3   &  5.95  \\
57.75  &0.76  &1.39 &  8.9   &  277.7  &  37.0   &  8.65   \\
60.0  & 0.74 &1.33 &  8.0   & 254.2   & 33.3    & 5.52    \\
62.5  &0.74  &1.39 &  8.0   &  259.9  &  33.3   &  9.55   \\
65.0  &0.72  &1.33 &  8.9   &  237.3  &  37.0   &  15.79   \\
70.7  & 0.715&1.39 & 12.0   & 238.8   &  49.9    &16.14  \\
 \hline\hline
\end{tabular}
\end{center}
\caption{The parameters and the volume integrals of the real and
imaginary potentials as well as the $\chi^{2}$ values of the Optical
model calculations. Coulomb radius, $r_{c}$=1.2 fm.}
 \label{param}
\end{table}
\newpage
\begin{table}[h]
\begin{center}
\begin{tabular}{lcccccc} \hline\hline
$E_{Lab}$ &$r_{V}$ &$a_{V}$&$W$& $J_{V}$ &$J_{W}$  &$\chi^{2}$\\
MeV & fm & fm & MeV  & MeV.fm$^{3}$ & MeV.fm$^{3}$ \\ \hline
 32.0  & 0.76  &1.30  &  3.2  &269.3  &13.3&   2.19    \\
 35.0  & 0.74  &1.37  &  3.2  &257.9  &13.3&   4.53    \\
 39.0  & 0.78  &1.38  &  3.2  &295.5  &13.3&   5.76    \\
 40.0  & 0.81  &1.39  &  3.5  &326.5  &14.5&   4.15    \\
 41.0  & 0.69  &1.45  &  4.2  &225.1  &17.4&   5.58    \\
 42.0  & 0.81  &1.35  &  4.9  &322.6  &20.4&   6.77    \\
 43.0  & 0.74  &1.42  &  5.3  &262.9  &22.0&   4.97    \\
 45.0  & 0.74  &1.45  &  5.2  &266.0  &21.6&   4.88    \\
 46.0  & 0.76  &1.37  &  5.6  & 275.7 &23.2&   6.92    \\
 49.0  & 0.80  &1.33  &  5.7  & 310.5 &23.7&   3.74    \\
 50.0  & 0.80  &1.32  &  5.8  & 309.6 &24.1&   3.64    \\
 52.0  & 0.739 & 1.39 &   7.7 &  259.0&32.0&    11.94  \\
 55.0  & 0.79  &1.37  &  8.0  & 304.3 & 33.2&  7.14   \\
 57.75  & 0.74  &1.45  &  8.3  &  266.0& 34.5&  8.49   \\
 60.0  & 0.77  &1.38  &  8.0  & 286.0 & 33.2&  6.32   \\
 62.5  & 0.72  &1.45  &  7.7  & 249.0 & 32.0&  8.36  \\
 65.0  & 0.70  &1.39  &  8.1  & 226.9 & 33.6&  11.86 \\
 70.7  & 0.77  &1.35   &12.0   &283.0 & 49.9& 13.59   \\
\hline\hline
\end{tabular}
\end{center}
\caption{The parameters and the volume integrals of the real and
imaginary potentials as well as the $\chi^{2}$ values of the
Coupled-Channels calculations. Coulomb radius, $r_{c}$=1.2fm.}
\label{paramcc}
\end{table}
\newpage
\begin{table}[h]
\begin{center}
\begin{tabular}{lcccccc} \hline\hline
$\theta_{C.M.}$ & $50^{0}$& $60^{0}$& $70^{0}$& $80^{0}$& $90^{0}$\\
OM (Eq. 2) & 40.16 & 4.44 & 26.82 & 67.15 & 31.96 \\
CC (Eq. 5) & 51.07 & 16.29 & 36.44 & 183.3 & 112.3\\
\hline\hline
\end{tabular}
\end{center}
\caption{$\chi^{2}$ values of the 50$^{0}$ to 90$^{0}$ excitation
functions by using the Optical and Coupled-Channels models.}
\label{chisqaure}
\end{table}

\newpage



\begin{thebibliography}{99}
\bibitem{Bro60} D.A. Bromley, J.A. Kuehner and E. Almquist, Phys. Rev. Lett. {\bf 4} (1960) 365.
\bibitem{Alm60} E. Almquist, D.A. Bromley and J.A. Kuehner and , Phys. Rev. Lett. {\bf 4} (1960) 515.
\bibitem{Erb81} K.A. Erb and D.A. Bromley,  Phys. Rev. C {\bf 23} (1981) 2781.
\bibitem{Cor77} T.M. Cormier, J. Applegate, G.M. Berkowitz, P. Braun-Munzinger, P.M. Cormier, J.W. Harris, C.M. Jachcinski, L. Lee, J. Barrette and H.E. Wegner, Phys. Rev. Lett. {\bf 38} (1977) 940.
\bibitem{Cor78} T.M. Cormier, C.M. Jachinski, G.M. Berkowitz, P. Braun-Munzinger, P.M. Cormier, M. Gai, J.W. Harris, J. Barrette and H.E. Wegner, Phys. Rev. Lett. {\bf 40} (1978) 924.
\bibitem{Ful80} B.R. Fulton, T.M. Cormier and B.J. Herman, Phys. Rev. {\bf C21} (1980) 198.
\bibitem{Cos75} E.R. Cosman, T.M. Cormier, K. Van Bibber, A. Sperduto, G. Young, J. Erskine, L.R. Greenwood and O. Hansen, Phys. Rev. Lett {\bf 35} (1975) 265.
\bibitem{Cos80} E.R. Cosman, R. Ledoux, and A.J. Lazzarini, Phys. Rev. C {\bf 21} (1980) 2111.
\bibitem{Sto77} R.G. Stokstad, R.M. Wieland, C.B. Fulmer, D.C. Hensley, S. Raman, A.H. Snell and P.H. Stelson, Oak Ridge National Laboratory, Report No. ORNL/TM-5935, 1977 (Unpublished).
\bibitem{Sto79} R.G. Stokstad, R.M. Wieland, G.R. Satchler, C.B. Fulmer, D.C. Hensley, S. Raman, L.D. Rickertsen, A.H. Snell and P.H. Stelson, Phys. Rev. C {\bf 20}
(1979) 655.
\bibitem{morsad} A. Morsad, F. Haas, C. Beck, and R.M. Freeman, Z. Phys. {\bf A338} (1979) 61.
\bibitem{Kolata} J.J.Kolata, R.M. Freeman, F. Haas, B. Heusch and A. Gallman, Phys. Rev. C 21 (1980) 579.
\bibitem{Bromley} K.A. Erb and D.A. Bromley, in Treatise on Heavy-Ion Science, Vol. 3, ed. D.A. Bromley
(Plenum, New York, 1985).
\bibitem{Gre1} W. Greiner, J.Y. Park, W. Scheid, Nuclear Molecules (World-Scientific, Singapore
1995).\\
G.R. Satchler, Direct Nuclear Reactions (Oxford Uni.
Press, Oxford 1983).\\
H. Feshbach, Theoretical Nuclear Physics: Nuclear Reactions (Wiley,
NewYork 1992).
\bibitem{Sat97} M.E. Brandan and G.R. Satchler, Phys. Rep. 285 (1997) 143.
\bibitem{freer04} M. Freer, M.P. Nicoli, S.M. Singer, C.A. Bremner, S.P.G. Chappell, W.D.M. Rae,
I. Boztosun, B.R. Fulton, D.L. Watson, B.J. Greenhalgh, G.K. Dillon,
R.L. Cowin and D.C. Weisser, Phys. Rev. C\textbf{70}(2004) 064311. \\
C.A. Bremner, S.P.G. Chappell, W.D.M. Rae, I. Boztosun, M. Freer,
M.P. Nicoli, S.M. Singer, B. Fulton, D.L. Watson, B.J. Greenhalgh,
G.K. Dillon, R.L. Cowin, Phys. Rev. C\textbf{66} (2002) 034605.
\bibitem{Mar86} S. Marsh, and W.D.M. Rae, Phys. Lett. {\bf
180B} (1986) 185.
\bibitem{Flo84} H. Flocard, P.H. Heenen, S.J. Krieger, and M.S. Weiss, Prog.
Theor. Phys. {\bf 72}(1984) 1000.
\bibitem{Ord86} C.E. Ordo\~{n}ez, R.J. Ledoux and E.R. Cosman, Phys. Lett. {\bf 173B} (1986) 39.
\bibitem{Kon79} Y. Kondo, Y. Abe and T. Matsuse, Phys. Rev. C {\bf 19} (1979) 1356.
\bibitem{Abe80} Y.Abe, Y. Kondo and T. Matsuse, Suppl. Prog. Theor. Phys. {\bf
68} (1980) 303.
\bibitem{Mat78} T. Matsuse, Y. Abe, Y. Kondo, Prog. Theor. Phys. {\bf 59} (1978) 1904.
\bibitem{Mor91} A. Morsad, F. Haas, C. Beck, and R.M. Freeman, Z. Phys {\bf
338} (1991) 61.
\bibitem{Row77} N. Rowley, H. Doubre, C. Marty, Phys. Letts. {\bf 69B} (1997) 147.
\bibitem{Bra90} M.E. Brandan, M. Rodriguez-Villafuerte and A. Ayala, Phys. Rev. C {\bf 41} (1990) 1520.
\bibitem{Kon98} Y. Kond\=o, M.E. Brandan and G.R. Satchler, Nucl. Phys. {\bf A637} (1998) 175.
\bibitem{mcvoy} K.W. McVoy and M.E. Brandan, Nucl. Phys. {\bf A542} (1992) 295.
\bibitem{Bra96} M.E. Brandan, M.S. Hussein, K.W. McVoy and G.R. Satchler, Comments Nucl. Part. Phys. 22 (1996) 77.
\bibitem{Boz1} I. Boztosun and W.D.M Rae, Phys. Rev. C {\bf 63} (2001) 054607.\\
I. Boztosun and W.D.M. Rae, Phys. Lett. {\bf 518B} (2001)229.
\bibitem{michel} F Michel and S. Ohkubo, Eur. Phys. J. {\bf A19} (2004) 333.
\bibitem{sat91} G.R. Satchler, Phys. Rep. 199 (1991) 147.
\bibitem{Sak84} Y. Sakuragi and M. Kamimura, Phys. Lett. {\bf 149B} (1984)
307. \\Y. Sakuragi, M. Yahiro and M. Kamimura, Prog. Theor. Phys.
\textbf{70} (1983) 1047.\\
M. Ito, Y. Sakuragi and Y. Hirabayashi, Phys. Rev. C {\bf 63},
064303 (2001).\\
Y. Sakuragi, M. Ito, M. Katsuma, M. Takashina, Y. Kudo, Y.
Hirabayashi, S. Okabe, and Y. Abe, in {\em Proceedings of the 7th
International Conference on Clustering Aspects of Nuclear Structure
and Dynamics}, edited by M. Korolija, Z. Basrak, and R. Caplar
(World Scientific, Singapore, 2000), p. 138.
\bibitem{Ste66} R.H. Stelson and L. Grodzins, Nucl. data {\bf 1A} (1966) 21. \\ F. Ajzenberg-Selove, Nucl. Phys. {\bf A248} (1975) 1.
\bibitem{chuck} P.D. Kunz, CHUCK, a Coupled-Channels code, unpublished.
\bibitem{fresco1} I.J. Thompson, FRESCO, a Coupled-Channels code, unpublished.\\
I.J. Thompson, Computer Physics Reports {\bf 7} (1988) 167.
\end{thebibliography}
\end{document}